# The Speed of Light and the Hubble parameter: The Mass-Boom Effect

*Antonio Alfonso-Faus*

*E.U.I.T. Aeronáutica*

*Plaza Cardenal Cisneros s/n*

*28040 Madrid, Spain*

ABSTRACT. We prove here that Newton's universal gravitation and momentum conservation laws together reproduce Weinberg's relation. It is shown that the Hubble parameter H must be built in this relation, or equivalently the age of the Universe t. Using a wave-to-particle interaction technique we then prove that the speed of light c decreases with cosmological time, and that c is proportional to the Hubble parameter H. We see the expansion of the Universe as an apparent effect due to the LAB value of the speed of light $c_o$ taken as constant. We present a generalized red shift law and find a predicted acceleration for photons that agrees well with the result from Pioneer 10/11 anomalous acceleration. We finally present a cosmological model coherent with the above results that we call the Mass-Boom. It has a linear increase of mass with time as a result of the speed of light linear decrease with time and the conservation of momentum. We obtain the baryonic mass parameter equal to the curvature parameter, $\Omega_m = \Omega_k$, so that the model is of the type of the Einstein static, closed, finite, spherical, unlimited, with zero cosmological constant. Neither dark matter nor dark energy is required by this model. With an initial constant speed of light during a short time we get inflation (an exponential expansion). This converts, during the inflation time, the Planck's fluctuation length of $10^{-33}$ cm to the present size of the Universe (about $10^{28}$ cm, constant from then on). Thereafter the Mass-Boom takes care to bring the initial values of the Universe (about $10^{15}$ gr) to the value at the present time of about $10^{55}$ gr.

1. - INTRODUCTION

In 1972 Weinberg [1] presented a relation that contained quantum parameters together with the Hubble cosmological parameter H. The order of magnitude of the mass of the fundamental particles was predicted and the relation is still catalogued as "unexplained". It has also been classified as a coincidence of numerical numbers and some explanations [2] have been proposed, none of them gaining general acceptance. Here we prove it using Newton's universal law of gravitation together with the momentum conservation law. We take a Machean approach considering the total momentum content of a mass M as Mc, a constant. The reality of the mass M, or the total momentum content Mc, is due to the interaction with the rest of the Universe. The total relativistic energy $Mc^2$ is also due to the interaction with the rest of the Universe. With these ideas in mind, together with a time varying speed of light as we will see, we derive the Weinberg relation as a consequence of Newton's laws.

By using a wave to particle interaction technique presented elsewhere, [3] and [4], we derive a generalized red shift law that contemplates time variations in c and G (the gravitational "constant"). Then we prove that the speed of light is proportional to the Hubble parameter, and therefore time-varying. This time variation explains the anomalous acceleration found in space probes like the Pioneer 10/11. With the propagation equation of particles and photons we prove two very important constancies: all gravitational radii are constant. In particular the gravitational radius of the Universe is constant. Also the cosmological scale factor of the Universe is constant: neither expansion nor contraction is present in the Universe at any scale.

The time variation of c implies a time variation of any mass. This is a result of the constancy of momentum: since the relation of any speed v to the speed of light c, v/c, has to be a constant to preserve special relativity, the constancy of momentum mv implies the constancy of mc. Then a decreasing speed of light implies an increasing mass, a Mass-Boom as we call it.

Finally we make use of the Einstein cosmological equations to find the cosmological model that this theory predicts: a closed, finite, unlimited, and curved (k = 1) Universe with zero cosmological constant and a baryon matter content given by $\Omega_m = \Omega_k$. Neither dark matter nor dark energy is necessary in this model. However, an initial inflationary phase is also built in the model that brings the initial Planck fluctuation to the present size of the Universe and the initial mass of $10^{15}$ gr.

2. – WEINBERG'S RELATION

In 1972 Weinberg [1] discussed a relation obtained by using G (gravitation), c (relativity), ℏ (quantum mechanics) and H (cosmology) to arrive at the mass m of a fundamental particle:

$$m^{1/3} \approx \frac{\hbar^2 H}{Gc} \quad (1)$$

We will follow a Machean approach by considering that the maximum momentum content of a mass M is Mc, a constant due to the rest of the Universe. It has acquired this momentum during a time t. Then the total force that can exert is Mc/t . The result of this heuristic argument can also be obtained by considering the rate of change with time of the total relativistic energy of the mass M, Mc², divided by c:

$$\frac{1}{c}\frac{dMc^2}{dt} = M\frac{dc}{dt} \qquad (2)$$

We later will prove that the speed of light c is proportional to the Hubble parameter H, c = HL, where L is a constant. For a linearly expanding Universe the time derivative of H is $-H^2$ so that in absolute value we have for (2)

$$M\frac{dc}{dt} = MH^2L = MHc \qquad (3)$$

Here we note that the Hubble parameter can be taken, in certain cases, as an operator that obtains the derivative, with respect to the cosmological time, of the function after it. To obtain the cosmological change with time we just put H in front of the function. This total force (3) can be thought of as fluxing over a spherical surface of radius r centred on the centre of gravity of M. It will act upon the small area represented by m, m << M, that has a size of the order of its Compton wavelength. Then m will feel a force exerted by the presence of M and given by

$$\frac{GMm}{r^2} = MHc\frac{(\hbar/mc)^2}{4\pi r^2} \qquad (4)$$

and rearranging we arrive at

$$4\pi m^3 = \frac{\hbar^2 H}{Gc} \qquad (5)$$

which is the Weinberg's relation (1) within a factor of less than 3.

3. – GENERALIZED RED SHIFT LAW

In 1983 Adams [3] presented a wave to particle technique to derive the matter and photon propagation. Later, Alfonso-Faus [4] applied this technique to the case of a time-varying gravitational constant G. Now we do the same but allowing the speed of light c to vary with cosmological time t too. Since the factor in the Einstein's field equations in front of the stress-energy tensor is $G/c^4$, we have now the generalized red shift law equation as

$$G/c^4 \, N_\gamma \, \hbar\nu R = \text{constante} \qquad (6)$$

Taking into account that the number of photons in a proper volume $N_\gamma$ is constant (no particle creation), and using $\nu = c/\lambda$, we get eliminating G from (5) and (6)

$$H \, N_\gamma \, R \, (\hbar/mc)^3 = c \, \lambda \qquad (7)$$

The wavelength λ is proportional to the cosmological scale factor R, and the Compton wavelength ℏ/mc is constant. Then we have

$$c = H \times \text{constant} = H \, L \qquad (8)$$

where L ≈ $10^{28}$ cm, is the present size of the visible Universe ct. Since the Hubble parameter H is inversely proportional to the cosmological time t so is the speed of light c. Substituting (8) into (1) we get the new Weinberg's relation

$$m^3 \approx \frac{\hbar^2}{GL} \qquad (9)$$

Here we see that we have obtained a classical relation, G, m, ℏ and L all constants, and no time dependence is explicit. But leaving G and m to vary with time we have for ℏ and L constant

$$Gm^3 = \text{constant} = \hbar^2/L \qquad (10)$$

This is what is observed in the lunar range experiments with laser. Now, if we demand that the field equations of Einstein´s general relativity continue to be derived from the action principle, we need to maintain constant the factors in front of the action integrals. This means

$$G/c^3 = \text{constant} \quad \text{and} \quad mc = \text{constant} \qquad (11)$$

And these relations satisfy Weinberg's in the form given in (9).

## 4. - EXPLANATION OF THE PIONEER 10/11 ANOMALOUS ACCELERATION

The fact that the speed of light varies with time as in (8) has enormous cosmological implications. We can get the photon acceleration as

$$dc/dt = L\, dH/dt \tag{12}$$

In the laboratory the speed of light is taken as constant $c_0$. Then we see a linear expansion of the Universe as $R = c_0 t$. Since H is R'/R we get for the photon acceleration a

$$a = dc/dt = -L\, H^2 = -Hc \tag{13}$$

Taking a range of values for H from 60 to 75 Km/sec/Mpc, the theoretical acceleration (13) for the photons is

$$a = -(7.13 \pm 0.3)\ \text{cm/seg}^2 \tag{14}$$

The observed anomalous acceleration in the Pioneer 10/11 case [5] is

$$a_p = -(8.74 \pm 1.33)\ \text{cm/seg}^2 \tag{15}$$

which is within the margin given in (14). From this point of view the observed acceleration is not satellite acceleration. It is in the photons.

## 5. - THE CONSTANCY OF GRAVITATIONAL RADII

With the wave to particle technique [4] we get now for the matter propagation equation

$$\left( (\frac{G}{c^2} NP^\mu)_{;\alpha} \right) P^\alpha = 0 \tag{16}$$

Using $P^\mu = mU^\mu$ we get

$$U^\mu_{;\alpha} U_\alpha + \frac{d\, \ln\left(\frac{G}{c^2} mN\right)}{dt} U^\mu U^0 = 0 \tag{17}$$

and contracting with $U_\mu$ we finally get

$$\frac{d\,\ln\left(\frac{G}{c^2}mN\right)}{dt}=0 \quad i.e.\quad \frac{GmN}{c^2}=const \tag{18}$$

This means that all the gravitational radii of masses mN inside its proper volume are constant. The possible expansion or contraction of the Universe does not affect them.

## 6. - THE CONSTANCY OF THE COSMOLOGICAL SCALE FACTOR

We have derived elsewhere [6] the expression for the zero value of the right hand side of the Einstein's field equations

$$\nabla\,(G/c^4\,.\,T^{\mu\nu}) = 0 \tag{19}$$

which is

$$\frac{\rho'}{\rho}+3(\omega+1)H+\frac{\Lambda'c^4}{8\pi G\rho}+\frac{G'}{G}-4\frac{c'}{c}=0 \tag{20}$$

where we have allowed for the time dependence of G, c and Λ. The cosmological parameter Λ is a real constant of integration so that it disappears from (20). In fact we will later show that Λ is zero. Here ρ is the energy density. Using the equation of state p = w ρ, integration of (20) gives

$$\frac{G\rho}{c^4} R^{3(\omega+1)} = const \quad (21)$$

or equivalently

$$GM/c^2 \cdot R^{3w} = \text{constant} \quad (22)$$

This result is very important. Since all the gravitational radii are constants and w is not zero, we arrive at the conclusion that the Universe does not expand nor contracts, i.e.

$$R = \text{constant} \quad (23)$$

Then the gravitational radius of the Universe can be equated to its visible size with R = ct = constant = L. For M the mass of the Universe at present we have

$$GM/c^2 = R = ct = L = \text{constant} \approx 10^{28} \text{ cm.} \quad (24)$$

No length in the Universe is expanding nor contracting. The dynamo paradox, that could ideally be constructed to obtain work from a rod attached to a galaxy that is going away from us, just goes away.

There is a way to interpret Mach's principle: the rest energy of any mass m is equal to its gravitational potential energy due to the rest of the mass of the Universe:

$$GMm/R = mc^2 \quad \text{or} \quad GM/c^2 = R \quad (25)$$

Comparing (24), (25) and (22) we get the value of w as

$$W = -1/3 \quad \text{so that} \quad p = -1/3\, \rho \quad (26)$$

We note that this is the same as the equation of state for photons, with a minus sign. We interpret it as the negative pressure of the gravity quanta.

## 7. - MASS-BOOM AND THE COSMOLOGICAL EQUATIONS

From (11) and (24) we get again the Mass-Boom, effect a linear increase of mass with cosmological time, M $\alpha$ t. We take the number of particles in the Universe N as constant so that in a certain system of units we have:

$$M = Nm = Nt \tag{27}$$

And from (24) we arrive at

$$G/c^3 = 1/N \quad \text{and} \quad m = t \tag{28}$$

The Einstein cosmological equations, with R = constant are then

$$8\pi \frac{Gp}{c^2} + \frac{Kc^2}{R^2} = \Lambda c^2$$
$$-\frac{8\pi}{3} G\rho_m + \frac{Kc^2}{R^2} = \frac{\Lambda c^2}{3} \tag{29}$$

Now with w = - 1/3 $\Lambda$ has to be identically zero. Then the two cosmological equations reduce to just one, i.e.

$$\frac{Kc^2}{R^2} = \frac{8\pi}{3} G \rho_m \tag{30}$$

We see that K = 1 is the solution to this equation. This cosmological model is spherical, closed, finite, unlimited, and static. Equation (30) can be interpreted as equilibrium between centrifugal forces, represented by the curvature term, and the negative pressure, the gravitational attraction.

In the usual dimensionless parameters $\Omega$ we have the equivalent to (30)

$$\Omega_m = \Omega_k \text{ (undetermined)} \tag{31}$$

We can take the value 0.05 for both parameters. Then neither dark matter nor dark energy is necessary in this model.

## 8. - THE INITIAL CONDITIONS OF THE UNIVERSE

During the first instants of time the speed of light must have been almost constant. This means that the Hubble parameter then was constant so that we have at that time:

$$H = R'/R = \text{constant} \tag{32}$$

Integrating the above relation we get the solution:

$$R = R_0 \, e^{Ht} \tag{33}$$

This exponential expansion is equivalent to the very well known inflation phase at the initial stages of the Universe. During the inflation phase the first Planck's fluctuation of size $10^{-33}$cm expanded very rapidly to the size $10^{28}$cm, as of today, so that we have

$$e^{Ht} = 10^{60} \tag{34}$$

If $t_1$ is the first tic of the Universe, and $t_i$ is the duration of the inflationary phase, then

$$t_i/t_1 = Ht = 60 \ln 10 \approx 138 \qquad (35)$$

Hence, only 138 tics were necessary to inflate the Planck's fluctuation to the present size of the Universe, and thereafter it remained at constant size. In our theory Planck's units have a time dependence as

$$\left(\frac{\hbar c}{G}\right)^{1/2} = 10^{60}/c = 10^{20} t \approx 10^{-5} \, gr$$

$$\left(\frac{G\hbar}{c^3}\right)^{1/2} = const = 1/10^{20} \approx 10^{-33} \, cm \qquad (36)$$

$$\left(\frac{G\hbar}{c^5}\right)^{1/2} = 1/(10^{20} c) = t/10^{60} \approx 10^{-44} \, s$$

These units had an initial value of

$$t = 10^{40}$$
$$t_1 = 1$$
$$\left(\frac{\hbar c}{G}\right)^{1/2}_1 = 10^{20} t_1 \approx 10^{-45} \, gr \qquad (37)$$
$$\left(\frac{G\hbar}{c^3}\right)^{1/2}_1 \approx 10^{-33} \, cm = cons\tan te$$
$$\left(\frac{G\hbar}{c^5}\right)^{1/2}_1 = t_1/10^{60} \approx 10^{-83} \, s$$

Here we see that inflation by a factor of $10^{60}$ gives an initial stage of the Universe as

$$M_i = 10^{15} \, gr., \quad L = 10^{28} \, cm = const., \quad t_i = 10^{-23} \, sec. \qquad (38)$$

From then on we have $5 \times 10^{40}$ tics to arrive at

$$M_0 = 5 \times 10^{55} \, gr., \quad L = 10^{28} \, cm = const., \quad t_0 = 5 \times 10^{17} \, sec. \qquad (39)$$

which is the present state of the Universe.

## 9. - PREDICTIONS FOR EXPERIMENTS

The theory presented here predicts that the Zeeman displacement varies with time as $1/t$. Then an experiment repeated two or three times every 3 to 5 years may appreciate this time variation. On the other hand the earth' gravitational acceleration varies as $1/t^2$. All of this means a time variation of the order of 1 part in $10^{10}$ per year. Within a period of five to ten years from now these variations may be detected.

## 10. - CONCLUSIONS

The wave-to-particle interaction technique is a powerful tool to determine very important relations in Nature. We have found the matter and photon propagation equations using the Einstein's general relativity field equations. We have found a generalized red shift law. We have allowed for time-varying G, m and c. We have found the new constants of Nature $G/c^3$, mc and c/H. We have defined a Mass-Boom cosmological model, a new frame of work for cosmology that has a decreasing speed of light and an increasing mass with cosmological time. The predicted photon acceleration agrees well with the anomalous acceleration of the Pioneer 10/11 satellites.

## 11. - REFERENCES